\documentclass[preprint2]{aastex6}
\usepackage{CJK}
\usepackage{color}

\usepackage{natbib}
\usepackage{hyperref}
\usepackage{cleveref}
\begin{document}
%\begin{CJK*}{GBK}{song}
\title{The key factor to determine the relation between radius and tidal deformability of neutron stars: slope of symmetry energy}
\author{Nai-Bo Zhang\altaffilmark{1}, Bin Qi\altaffilmark{1}, and Shou-Yu Wang$^{*}$\altaffilmark{1}}
\altaffiltext{1}{Shandong Provincial Key Laboratory of Optical Astronomy and Solar-Terrestrial Environment,
Institute of Space Sciences, Shandong University, Weihai, 264209, China}
\noindent{$^{*}$Corresponding author: sywang@sdu.edu.cn}
\begin{abstract}
The constraints on tidal deformability $\Lambda$ of neutron stars are first extracted from GW170817 by LIGO and Virgo Collaborations but the relation between radius $R$ and tidal deformability $\Lambda$ is still nuder debate. Using an isospin-dependent parameterized equation of state (EOS), we study the relation between $R$ and $\Lambda$ of neutron stars and its dependence on parameters of symmetry energy $E_{\rm sym}$ and EOS of symmetric nuclear matter $E_0$ when the mass is fixed as $1.4$ $M_\odot$, $1.0$ $M_\odot$, and $1.8$ $M_\odot$, respectively. We find that, though the changes of high order parameters of $E_{\rm sym}$ and $E_0$ can shift the individual values of $R_{1.4}$ and $\Lambda_{1.4}$ to different values, the $R_{1.4}\sim\Lambda_{1.4}$ relation approximately locates at the same fitted curve. The slope of symmetry energy $L$ plays the dominated role in determining the $R_{1.4}\sim\Lambda_{1.4}$ relation. By checking the mass dependence of $R\sim\Lambda$ relation, the well fitted $R\sim\Lambda$ relation for 1.4 $M_\odot$ is broken for massive neutron stars.
\end{abstract}
\keywords{Dense matter, equation of state, stars: neutron}
\maketitle
%\newpage

\section{Introduction}

The equation of state (EOS) of dense neutron-rich matter plays the dominated role in determining the properties of neutron stars. In turn, the observations of neutron stars can put strict constraints on the EOS. Currently, the most widely used observation to constrain the EOSs is the maximum observed mass of pulsars J1614-2230 with mass $M=1.908\pm0.016$ $M_\odot$ \citep{Demorest10,Arzoumanian2018} and J0348+0432 with $M=2.01\pm0.04$ $M_\odot$ \citep{Antoniadis13}. Many soft EOSs that can not support such high mass have been excluded. Recently, the newly reported pulsar J0740+6620 with $M=2.17^{+0.11}_{-0.10}$ $M_\odot$ \citep{Cromartie19} induces a new theoretical challenge in satisfying the constrained pressure band for several typical microscopic nuclear EOSs with maximum mass $M_{\rm max}>2.17$ $M_\odot$ \citep{Zhang19c}.

Except the mass, the radius of neutron stars is another observable to constrain the EOS. To extract the radii, special attentions have been focusing on the quiescent low-mass X-ray binaries \citep{Campana04,Heinke04} or the isolated cooling neutron stars \citep{Mereghetti11} and various analyses have been performed \citep{Guillot13,Steiner13,Guillot14,Lattimer14,Bogdanov16,Guillot16,Ozel16,Nattla17,Shaw18,Steiner18}. However, due to the great difficulties in, such as, determining the precise distance of the source and compositions of atmosphere, large uncertainties still exist and additional information is required to estimate the correct radii (see, e.g., \citet{Miller13,Ozel13,Lattimer14,Miller16,Suleimanov16} for recent reviews).

The first joint gravitational and electromagnetic observation of GW170817 opens the era of multimessenger astronomy \citep{LIGO17,LIGO17b} and further motivates abundant studies on the EOS in both astrophysics and nuclear physics communities (see, e.g., \citet{Piekarewicz18b,Li19,Raithel19} for recent reviews). In a coalescing binary neutron star system, one neutron star suffers the tidal deformation induced by the strong tidal field of its companion. The dimensionless tidal deformability of canonical neutron stars was first extracted \citep{LIGO17} and then refined as $\Lambda_{1.4}=190^{+390}_{-120}$ at 90\% confidence level by assuming that both neutron stars are described by the same EOS and spin at low-spin prior \citep{LIGO18}. Thereafter, various constraints have been put on the properties of nuclei and neutron stars, such as, the neutron-skin thickness of $^{208}$Pb \citep{Fattoyev18,Nandi18,Tong19}, the radius $R$ \citep{LIGO18,Nandi18,Lourenco19,Raithel18,Fattoyev18,Malik18,Zhou19,Most18,Annala18,Lim18,Radice19,Tews18,De18,Bauswein17}, and maximum mass $M_{\rm max}$ \citep{Margalit17,Shibata17,Rezzolla18,Ruiz18,Zhou18,Baym19,Shibata19,Zhang19} of neutron stars.

As the tidal deformability $\Lambda$ is related to radius $R$ by $\Lambda = 2k_2/3(c^2R/GM)^5$ \citep{Hinderer08,Hinderer10} where $k_2$ is the second Love number, an underlying relation may exist between $\Lambda$ and $R$ when the mass is fixed. If the relation is explicit, one quantity can be extracted when another is measured/determined. However, as the $k_2$ has to be solved by a complicated differential equation coupled to TOV equations \citep{Tolman34,Oppenheimer39}, the exact relation between radius and tidal deformability of canonical neutron stars ($R_{1.4}\sim\Lambda_{1.4}$ relation) is still not well determined. In the present work, we focus on three sides related to the relation between radius and tidal deformability: (1) Calculate the $R_{1.4}\sim\Lambda_{1.4}$ relation based on an isospin-dependent parameterized EOS; (2) Delineating the dependence of $R_{1.4}\sim\Lambda_{1.4}$ relation on the parameters of nuclear matter; (3) Revealing the mass dependence of $R\sim\Lambda$ relation.

\begin{figure}[htb]
\begin{center}
  \includegraphics[width=8cm]{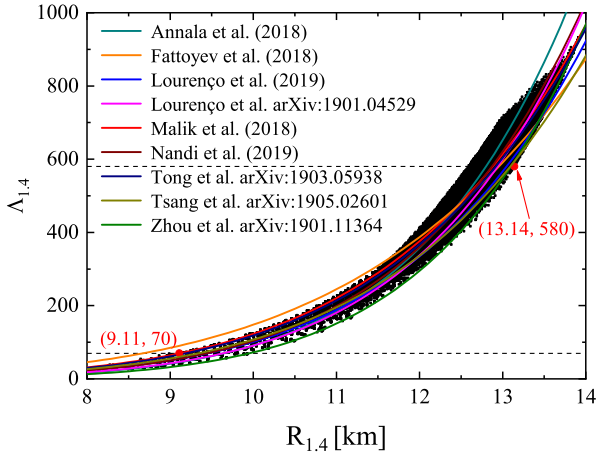}
  \caption{(color online) The tidal deformability $\Lambda_{1.4}$ as a function of radius $R_{1.4}$ of canonical neutron stars based on the isospin-dependent parameterized EOS. The parameters $L$, $K_{\rm sym}$, $J_{\rm sym}$, and $J_0$ are valued in steps  of 5, 50, 100, and 50 MeV within their uncertainties, respectively. The horizontal dashed lines represent the refined constraints of $70<\Lambda_{1.4}<580$ (90\% confidence level) extracted by LIGO and Virgo Collaborations~\citep{LIGO18}. The red dots labeled by the coordinates represent the constrained upper and lower limit on $R_{1.4}$. The solid lines correspond to the fitted equations suggested by different work. }\label{LRtotal}
\end{center}
\end{figure}

The $R_{1.4}\sim\Lambda_{1.4}$ relation has been discussed in \citet{Nandi18,Lourenco19,Fattoyev18,Malik18,Zhou19,Lourenco18,Tsang19,Annala18,Kim18,Lim18,Annala18} based on the EOSs calculated from relativistic mean field (RMF) theory, Skyrme Hartree-Fock (SHF) theory, microscopic theories, or parameterized EOSs. Various $R_{1.4}\sim\Lambda_{1.4}$ relations have been fitted and summarized in Figure \ref{LRtotal}. We can see that, though all the fitted equations spread within a narrow band in the $R_{1.4}\sim\Lambda_{1.4}$ plane, the uncertainty still remains.

The radius of neutron stars is known to be sensitive to the symmetry energy around 2 times saturation density \citep{Lattimer00,Lattimer01}. The dependence of $\Lambda_{1.4}$ on symmetry energy was first discussed in \citet{Fattoyev13,Fattoyev14} where they concluded that the tidal deformability of low mass neutron stars is only sensitive to the slope $L$ of symmetry energy while the tidal deformability of massive neutron stars is sensitive to the high density behavior of symmetry energy. Recently, \citet{Tsang19} studied the relation between $\Lambda_{1.4}$ and the parameters of symmetry energy $E_{\rm sym}$ and EOS of symmetric nuclear matter (SNM) $E_0$ based on more than 200 Skyrme EOSs and found that strong relation exists between $\Lambda_{1.4}$ and the curvature of symmetry energy $K_{\rm sym}$. In addition, in several of our previous work \citep{Zhang18,Zhang19,Zhang19b}, we have shown the individual results of $R_{1.4}$ and $\Lambda_{1.4}$ in the 3 dimensional parameter space of an isospin-dependent parameterized EOS characterized by the parameterizations of $E_{\rm sym}$ and $E_0$. It is shown that the slope $L$ and curvature $K_{\rm sym}$ of symmetry energy play almost the equally important role in determining the individual values of $R_{1.4}$ and $\Lambda_{1.4}$. In the present work, instead of the individual results of $R_{1.4}$ and $\Lambda_{1.4}$, the dependence of $R_{1.4}\sim\Lambda_{1.4}$ relation on the $E_{\rm sym}$ or $E_0$ is discussed systematically and the slope $L$ of symmetry energy is found to play the dominated role in determining the $R_{1.4}\sim\Lambda_{1.4}$ relation.

Besides, the $R\sim\Lambda$ relation is normally studied for the neutron stars with $M=1.4$ $M_\odot$. Whether the $R\sim\Lambda$ relation still holds for low mass or massive neutron stars (such as, $R_{1.0}\sim\Lambda_{1.0}$ relation or $R_{1.8}\sim\Lambda_{1.8}$ relation) needs further studies. The $R\sim\Lambda$ relation for a given EOS has been calculated and shown in, e.g., \citet{Piekarewicz18,Zhang19b}). It is found that the $R$ first increases and then stays approximately a constant for increasing $\Lambda$ (decreasing $M$). In the present work, the mass dependence (fixing mass) of $R\sim\Lambda$ relation is studied and it is found that the well fitted $R\sim\Lambda$ relation for 1.4 $M_\odot$ is broken for massive neutron stars.

Based on the above discussions, we study the relation between radius and tidal deformability of neutron stars. Specifically, we study the $R\sim\Lambda$ relation and its dependence on the symmetry energy and EOS of SNM when the mass is fixed as $1.4$ $M_\odot$, $1.0$ $M_\odot$, and $1.8$ $M_\odot$, respectively. We find that the slope $L$ of symmetry energy plays the dominated role in determining the $R_{1.4}\sim\Lambda_{1.4}$ relation and the well fitted $R\sim\Lambda$ relation for 1.4 $M_\odot$ is broken for massive neutron stars. The theoretical framework is summarized in Section \ref{sec2}. The discussions of $R_{1.4}\sim\Lambda_{1.4}$ relation and its dependence on $E_{\rm sym}$, $E_0$, and mass is shown in Section \ref{sec3}, and the conclusions are given in Section \ref{sec4}.

\section{Theoretical Framework}
\label{sec2}

Based on the definitions of the parameters of symmetry energy $E_{\rm sym}$, namely, the slope $L=3\rho_0[\partial E_{\rm sym}(\rho)/\partial\rho]_{\rho=\rho_0}$, the curvature $K_{\rm sym}=9\rho_0^2[\partial^2 E_{\rm sym}(\rho)/\partial\rho^2]_{\rho=\rho_0}$, and the skewness $J_{\rm sym}=27\rho_0^3[\partial^3 E_{\rm sym}(\rho)/\partial\rho^3]_{\rho=\rho_0}$, these parameters are potentially correlated for the EOSs calculated from RMF, SHF, or other theories, which hampers the applications of studying their independent role in determining the properties of neutron stars. In addition, the calculated relations among mass, radius, and tidal deformability are dependent on the chosen EOSs and cannot change continuously. On the contrary, the parameterized EOSs allow us to study the properties of neutron stars systematically by varying the parameters continuously though the physics behind the EOSs is partly missing.

Recently, \citet{Zhang18} constructed an isospin-dependent parameterized EOS describing the neutron stars with core consisting of neutron, proton, electron, and muon at $\beta$-equilibrium (charge neutral $npe\mu$ matter) and developed a numerical technique of inverting the TOV equations to constrain the EOS and symmetry energy based on the observations of radius, tidal deformability, and maximum mass, or physical requirement, such as, the causality condition (speed of sound is smaller than speed of light) \citep{Zhang19,Zhang19b,Zhang19c}. The construction and demonstration have been presented in detail in \cite{Zhang18}. For completeness and easy of discussions, we first summarize here the key aspects of the model. As an EOS is mainly determined by the EOS of asymmetric nucleonic matter $E_b(\rho,\delta)$ with isospin asymmetry $\delta=(\rho_{\rm{n}}-\rho_{\rm{p}})/\rho$ at density $\rho$ through $\epsilon(\rho, \delta)=\rho [E_b(\rho,\delta)+M_N]+\epsilon_l(\rho, \delta)$ and $E_b(\rho,\delta)$ can be approximated as \citep{Bom91}
\begin{equation}\label{PAEb}
  E_b(\rho,\delta)=E_0(\rho)+E_{\rm{sym}}(\rho)\delta^2,
\end{equation}
we parameterize $E_0(\rho)$ and $E_{\rm sym}(\rho)$ as
\begin{eqnarray}\label{E0para}
  E_{0}(\rho)&=&E_0(\rho_0)+\frac{K_0}{2}(\frac{\rho-\rho_0}{3\rho_0})^2+\frac{J_0}{6}(\frac{\rho-\rho_0}{3\rho_0})^3,\\
  E_{\rm{sym}}(\rho)&=&E_{\rm{sym}}(\rho_0)+L(\frac{\rho-\rho_0}{3\rho_0})+\frac{K_{\rm{sym}}}{2}(\frac{\rho-\rho_0}{3\rho_0})^2\nonumber\\
  &+&\frac{J_{\rm{sym}}}{6}(\frac{\rho-\rho_0}{3\rho_0})^3,\label{Esympara}
\end{eqnarray}
where $\rho_0$ is the saturation density of SNM.

The above equations have the same form as the widely used Taylor expansions around $\rho_0$. With the density increasing, the convergence problem appears for Taylor expansions. However, as demonstrated in great detail in \citet{Zhang18}, the above equations can still be used to simulate nuclear matter at high density if we see the coefficients as free parameters that should be determined by the observations and the parameterizations naturally become the Taylor expansions when $\rho\rightarrow \rho_0$. As free parameters, in principle, they can be varied as any values. In this case, it is infeasible to perform a study with 7 free parameters. To reduce the freedom of parameters, we first fixed the well constrained parameters as their currently known most probable values around $\rho_0$ extracted from terrestrial nuclear laboratory experiments, namely, $E_0(\rho_0)=-15.9$ MeV \citep{Brown14}, $E_{\rm{sym}}(\rho_0)=-31.7$ MeV \citep{Li13,Oertel17,Li17}, and $K_0=230$ MeV \citep{Shlomo06,Piekarewicz10}. As the uncertainty of $L$ is constrained as $\approx 58.7\pm 28.1 $ MeV \citep{Li13,Oertel17,Li17} but $J_0$, $K_{\rm sym}$, and $J_{\rm sym}$ are leses constrained ($-800\leq J_0\succ400$ MeV, $-400\leq K_{\rm sym}\leq100$ MeV, and $-200\leq J_{\rm sym}\leq800$ MeV) \citep{Tews17,Zhang17}, we then fixed $L=58.7$ MeV and studied how the observations of maximum observed mass $M_{\rm max}$, radius $R_{1.4}$ and tidal deformability $\Lambda_{1.4}$ of canonical neutron stars, and also causality condition can constrain the EOS and symmetry energy in the 3 dimensional $K_{\rm sym}-J_{\rm sym}-J_0$ parameter space \citep{Zhang19}. We found that $J_0$ and thereby $E_0$ have slight effects on the radius and tidal deformability of canonical neutron stars. Then, the effects of symmetry energy on the properties of neutron stars were delineated in the 3 dimensional $L-K_{\rm sym}-J_{\rm sym}$ parameter space \citep{Zhang19b}. In the present work, the overall effects of $L-K_{\rm sym}-J_{\rm sym}-J_0$ parameters on $R_{1.4}$, $\Lambda_{1.4}$, and also $R_{1.4}\sim\Lambda_{1.4}$ relation are discussed.

Using the parameterized EOS to describe the core of neutron stars, the outer and inner crusts are replaced by BPS \citep{Baym71} and NV EOSs \citep{Negele73} where the transition density is self-consistently calculated by investigating the incompressibility \citep{Kubis04,Kubis07,Lattimer07,Xu09}:
\begin{eqnarray}\label{tPA}
K_\mu&=&\rho^2\frac{d^2E_0}{d\rho^2}+2\rho\frac{dE_0}{d\rho}\\ \nonumber
&+&\delta^2\left[\rho^2\frac{d^2E_{\rm{sym}}}{d\rho^2}+2\rho\frac{dE_{\rm{sym}}}{d\rho}-2E^{-1}_{\rm{sym}}(\rho\frac{dE_{\rm{sym}}}{d\rho})^2\right].
\end{eqnarray}
Once the $K_\mu$ becomes negative, the thermodynamical instability grows by forming clusters, indicating a transition from the uniform core to the clustered crust. Note that, as discussed in \citet{Piekarewicz18,Gamba19}, the crust EOSs have little effects on the maximum mass and tidal deformability.

For a given EOS, the mass and radius can be calculated by solving the TOV equations \citep{Tolman34,Oppenheimer39}:
\begin{equation}\label{TOVp}
\frac{dP}{dr}=-\frac{G(m(r)+4\pi r^3P/c^2)(\epsilon+P/c^2)}{r(r-2Gm(r)/c^2)},
\end{equation}
\begin{equation}\label{TOVm}
\frac{dm(r)}{dr}=4\pi\epsilon r^2,
\end{equation}
and the tidal deformability can be obtained by solving a complicated differential equation coupled to TOV equations \citep{Hinderer08,Hinderer10}.

\section{Relation between radius and tidal deformability of neutron stars}
\label{sec3}

As discussed in the introduction, in several of our previous work \citep{Zhang18,Zhang19,Zhang19b}, we have shown the individual results of $R_{1.4}$ and $\Lambda_{1.4}$ in the 3 dimensional parameter space of $K_{\rm sym}-J_{\rm sym}-J_0$ fixing $L=58.7$ MeV \citep{Zhang18,Zhang19} and $L-K_{\rm sym}-J_{\rm sym}$ fixing $J_0=-180$ MeV \citep{Zhang19b}. To further clarify the  $R_{1.4}\sim\Lambda_{1.4}$ relation and its dependence on the symmetry energy and EOS of SNM, the tidal deformability $\Lambda_{1.4}$ as a function of radius $R_{1.4}$ of canonical neutron stars based on the isospin-dependent parameterized EOS is shown in Figure \ref{LRtotal}. The parameters $L$, $K_{\rm sym}$, $J_{\rm sym}$, and $J_0$ are valued in steps  of 5, 50, 100, and 50 MeV within their uncertainties, respectively. The horizontal dashed lines represent the refined constraints of $70<\Lambda_{1.4}<580$ (90\% confidence level) extracted by LIGO and Virgo Collaborations~\citep{LIGO18}. The red dots labeled by the coordinates represent the constrained upper and lower limits of $R_{1.4}$. The solid lines correspond to the fitted equations as discussed in the introduction. We can see that almost all the fitted equations can be covered by the present calculations of parameterized EOS, except the one from \citet{Fattoyev18} where only EOSs with $R_{1.4}>12.5$ km are adopted and the $R_{1.4}\sim\Lambda_{1.4}$ relation is extrapolated with larger errors for smaller radii. Thus, the parameterized EOSs are sufficient to study the $R_{1.4}\sim\Lambda_{1.4}$ relation systematically. If considering the refined constraint of tidal deformability, as labeled by red dots, the radius can be constrained as $9.11<R_{1.4}<13.14$ km. This upper limit is in great consistent with \citet{Annala18} using EOSs interpolating between chiral effective theory at low density and perturbative quantum chromo dynamics at high baryon density, and \citet{Li06} using available terrestrial laboratory data on the isospin diffusion in heavy-ion reactions at intermediate energies. If considering the constraint of $2.01$ $M_\odot$ from \citet{Antoniadis13} or the newly reported $2.17$ $M_\odot$ from \citet{Cromartie19}, more parameter sets with small radii can be excluded and thus rise the lower limit of $R_{1.4}$ apparently, as discussed in \citet{Annala18}. It should be noted that all parameterized EOSs are generated with the same confidence level and the data density is an artificial instead of physical results and dependent on the chosen parameter sets.

\begin{figure}[htb]
\begin{center}
  \includegraphics[width=8cm]{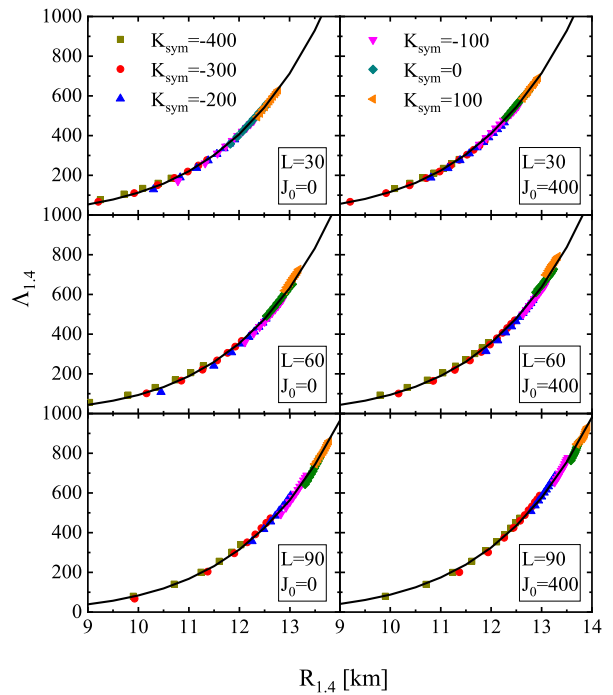}
  \caption{(color online) The tidal deformability $\Lambda_{1.4}$ as a function of radius $R_{1.4}$ of canonical neutron stars when $L$ ($J_0$) is fixed as 30, 60, and 90 MeV (0 and 400 MeV), respectively. The $J_{\rm sym}$ is varied from -200 to 800 MeV when $K_{\rm sym}$ is fixed as -400, -300, -200, -100, 0, and 100 MeV for each plot. The solid lines correspond to the fitted curves. }\label{LRKsymJsym}
\end{center}
\end{figure}

As the $E_0(\rho)$ and $E_{\rm sym}(\rho)$ have been parameterized in our calculations, except showing the overall $R_{1.4}\sim\Lambda_{1.4}$ relation, the individual effects of each parameter of $E_0(\rho)$ and $E_{\rm sym}(\rho)$ can also be studied. To study the effects of $K_{\rm sym}$ and $J_{\rm sym}$ parameters, the tidal deformability $\Lambda_{1.4}$ as a function of radius $R_{1.4}$ of canonical neutron stars when $L$ ($J_0$) is fixed as 30, 60, and 90 MeV (0 and 400 MeV) is shown in Figure \ref{LRKsymJsym}. The $J_{\rm sym}$ increases from -200 to 800 MeV when $K_{\rm sym}$ is fixed as -400, -300, -200, -100, 0, and 100 MeV for each panel. The solid lines correspond to the fitted curves in the form of $\Lambda_{1.4}=aR_{1.4}^b$. Note that $J_0$ has been constrained as $-220\sim200$ MeV in \citet{Zhang19} and $J_0=400$ MeV is adopted here to magnify the effects of $J_0$. We can see that the increase of $K_{\rm sym}$, $J_{\rm sym}$, or $J_0$ can shift the data to larger $\Lambda_{1.4}$ and $R_{1.4}$ along the fitted curves, which is independent of the chosen $L$. Using the fitted curve of $L=30$ MeV and $J_0=0$ MeV as a reference, the deviation between the data and fitted curve by varying $K_{\rm sym}$ and $J_{\rm sym}$ (top left panel) is less than 5\% except several points with small radii. If considering the effects of $L$, i.e., varying $L$ from 30 MeV to 90 MeV (bottom left panel), the deviation increases to about 20\%. Further, additional 2\% deviation is added if including the freedom of $J_0$ (bottom right panel).

\begin{figure}[htb]
\begin{center}
  \includegraphics[width=8cm]{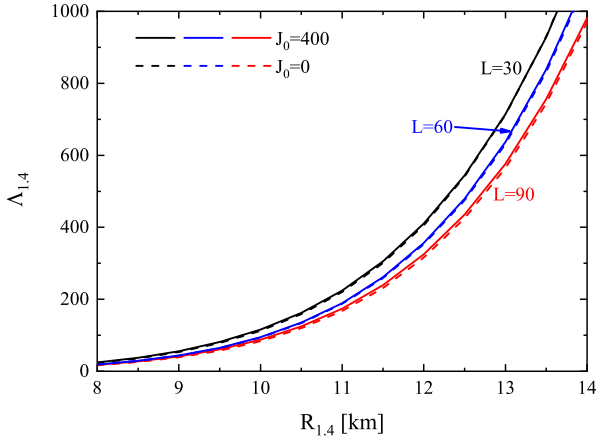}
  \caption{(color online) The fitted curves for different $L$ and $J_0$ parameter sets given in Figure \ref{LRKsymJsym} with $K_{\rm sym}$ and $J_{\rm sym}$ varying within their uncertainties. The black, blue, and red lines correspond to the results of $L=30$, $60$, and $90$ MeV. The solid and dashed lines correspond to results of $J_0=400$ and $0$ MeV.}\label{Leffect}
\end{center}
\end{figure}

To shown the above discussions clearly, we summarize the fitted curves in Figure \ref{Leffect}. The black, blue, and red lines correspond to the results of $L=30$, $60$, and $90$ MeV. The solid and dashed lines correspond to results of $J_0=400$ and $0$ MeV. We can see that, though $J_0$ changes from 400 to 0 MeV, the $R_{1.4}\sim\Lambda_{1.4}$ relation is nearly unchanged (corresponding to the 2\% deviation above). However, when $L$ changes from 30 to 90 MeV, the $R_{1.4}\sim\Lambda_{1.4}$ relation changes apparently (corresponding to the 20\% deviation above). Combining the discussions above, the slope of symmetry energy $L$ shows the dominated role in determining the $R_{1.4}\sim\Lambda_{1.4}$ relation though $L$ and $K_{\rm sym}$ play almost the equally important role in determining the individual values of $R_{1.4}$ and $\Lambda_{1.4}$ \citep{Zhang19b}. Thus, the precise measurement of $L$ is crucial to determine the exact relation between $R_{1.4}$ and $\Lambda_{1.4}$ which are excepted to be individually measured by the LIGO and Virgo Collaborations and Neutron star Interior Composition ExploreR (NICER), respectively. In fact, the effects of $J_0$ have been showed in \citet{Zhang18,Zhang19}. The constant surfaces of $R_{1.4}$ and $\Lambda_{1.4}$ are almost parallel to the $J_0$ axis and thus parallel to each other in the $K_{\rm sym}-J_{\rm sym}-J_0$ parameter space for $L=58.7$ MeV, which shows the weak dependence of $R_{1.4}$, $\Lambda_{1.4}$, and thereby $R_{1.4}\sim\Lambda_{1.4}$ relation on $J_0$.

\begin{figure}[htb]
\begin{center}
  \includegraphics[width=8cm]{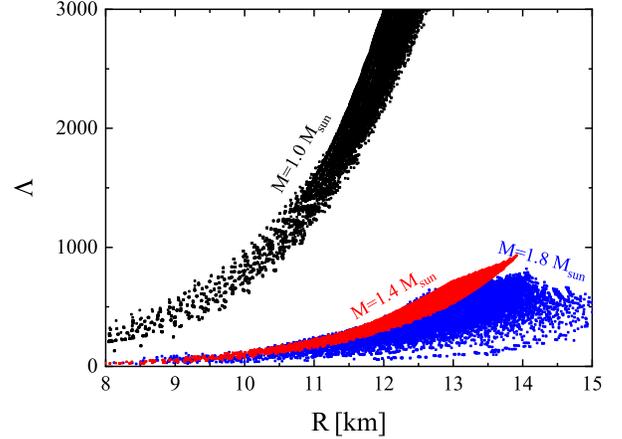}
  \caption{(color online) The tidal deformability as a function of radius for neutron stars with $1.0$ $M_\odot$ (black dots), $1.4$ $M_\odot$ (red dots), and $1.8$ $M_\odot$ (blue dots) based on the isospin-dependent parameterized EOS.}\label{LR1018}
\end{center}
\end{figure}

Universal relation appears when varying $K_{\rm sym}$, $J_{\rm sym}$, and $J_0$ for $M=1.4$ $M_\odot$. Whether the relation still survives for neutron stars with smaller or larger mass? To check the mass dependence of the above discussions, we calculate the tidal deformability as a function of radius for neutron stars with $1.0$ $M_\odot$ (black dots), $1.4$ $M_\odot$ (red dots), and $1.8$ $M_\odot$ (blue dots) and show the results in Figure \ref{LR1018}. We can see that, compared to the results of neutron stars with $1.4$ $M_\odot$, larger ranges can be covered for neutron stars with $1.0$ $M_\odot$. This is easy to understand as the $\Lambda$ decreases with increasing $M$ for a given EOS \citep[see, e.g.,][]{Zhang19b}. In addition, as the central density of neutron stars with $1.8$ $M_\odot$ is much larger than the stars with $1.4$ $M_\odot$, the high order parameters in Eqs. (\ref{E0para}) and (\ref{Esympara}) start to play more important roles in determining the properties of neutron stars and thus break the $R\sim\Lambda$ relation of canonical neutron stars.

\section{Conclusions}
\label{sec4}

Using an isospin-dependent parameterized equation of state (EOS), we study the relation between $R$ and $\Lambda$ of neutron stars and its dependence on symmetry energy $E_{\rm sym}$ and EOS of SNM $E_0$ when the mass is fixed as $1.4$ $M_\odot$, $1.0$ $M_\odot$, and $1.8$ $M_\odot$, respectively. We find that, though the changes of high order parameters of $E_{\rm sym}$ and $E_0$ can shift the individual values of $R_{1.4}$ and $\Lambda_{1.4}$ to different values, the $R_{1.4}\sim\Lambda_{1.4}$ relation approximately locates at the same fitted curve. The slope of symmetry energy $L$ plays the dominated role in determining the $R_{1.4}\sim\Lambda_{1.4}$ relation and the precise measurement of $L$ is crucial to determine the exact relation between $R_{1.4}$ and $\Lambda_{1.4}$. Compared to the results of neutron stars with $1.4$ $M_\odot$, larger ranges in the $R\sim\Lambda$ plane can be covered for neutron stars with $1.0$ $M_\odot$ and $1.8$ $M_\odot$. The well fitted $R\sim\Lambda$ relation for 1.4 $M_\odot$ is broken for massive neutron stars.

The possible relation among $L-R_{1.4}-\Lambda_{1.4}$ is found in the present work. All three of them remain inconclusive so far. Fortunately, any progress from heavy-ion reactions, measurement of radius (such as, NICER), or gravitational wave events (such as, LIGO and Virgo Collaborations) may help to break the degeneracy and constrain other two quantities.

\acknowledgments

We would like to thank Prof. Bao-An Li for constructive suggestions and comments. This work is partly supported by the Shandong Natural Science Foundation under Grants No. JQ201701, the Natural Science Foundation of China under Grants No. 11622540, No. 11675094, No. 11705102, No. 11775133, the China Postdoctoral Science Foundation under Grant No. 2019M652358, the Young Scholars Program of Shandong University, Weihai under Grant No. 2015WHWLJH01, and the Fundamental Research Funds of Shandong University under Grant No. 2019ZRJC001.


\begin{thebibliography}{}
\bibitem[Abbott et al.(2017)]{LIGO17} Abbott, B. P., et al. 2017, \prl, 119, 161101
\bibitem[Abbott et al.(2017b)]{LIGO17b} Abbott, B. P., et al. 2017b, \apjl, 848, L12
\bibitem[Abbott et al.(2018)]{LIGO18} Abbott, B. P., et al. 2018, \prl, 121, 161101
\bibitem[Annala et al.(2018)]{Annala18} Annala, E., Gorda, T., Kurkela, A., \& Vuorinen, A. 2018, \prl, 120, 172703
\bibitem[Antoniadis et al.(2013)]{Antoniadis13} Antoniadis, J., et al. 2013, Sci, 340, 448
\bibitem[Arzoumanian et al.(2018)]{Arzoumanian2018} Arzoumanian, Z., et al. 2018, \apjs, 235, 37
\bibitem[Bauswein et al.(2018)]{Bauswein17} Bauswein, A., Just, O., Janka, H., \& Stergioulas, N. 2017, \apjl, 850, L34
\bibitem[Baym et al.(1971)]{Baym71} Baym, G., Pethick, C. J., \& Sutherland, P. 1971, \apj, 170, 299
\bibitem[Baym et al.(2019)]{Baym19} Baym, G., Furusawa, S., Hatsuda, T., Kojo, T., \& Togashi, H. 2019, arXiv:1903.08963
\bibitem[Bogdanov et al.(2016)]{Bogdanov16} Bogdanov, S., Heinke, C. O., \"{O}zel, F., \& G\"{u}ver, T. 2016, \apj, 831, 184
\bibitem[Bombaci \& Lombardo(1991)]{Bom91} Bombaci, I., \& Lombardo, U. 1991, \prc, 44, 1892
\bibitem[Brown \& Schwenk(2014)]{Brown14} Brown, B. A., \& Schwenk, A. 2014, \prc, 89, 011307, Erratum: [2015, \prc, 91, 049902].
\bibitem[Campana \& Stella(2004)]{Campana04} Campana, S., \& Stella, L. 2004, Nucl. Phys. B Proc. Suppl., 132, 427
\bibitem[Cromartie {et al.}(2019)]{Cromartie19} Cromartie, H. T., et al. 2019, arXiv:1904.06759
\bibitem[De et al.(2018)]{De18} De, S., Finstad, D., Lattimer, J. M., Brown, D. A., Berger, E., \& Biwer, C. M. 2018, \prl, 121, 091102
\bibitem[Demorest et al.(2010)]{Demorest10} Demorest, P. B., Pennucci, T., Ransom, S. M., Roberts, M. S. E., \& Hessels, J. W. T. 2010, Natur, 467, 1081
\bibitem[Fattoyev et al.(2013)]{Fattoyev13} Fattoyev, F. J., Carvajal, J., Newton, W. G., \& Li, B. A. 2013, \prc, 87, 015806
\bibitem[Fattoyev et al.(2014)]{Fattoyev14} Fattoyev, F. J., Newton, W. G., \& Li, B. A. 2014, Eur. Phys. J. A, 50, 45
\bibitem[Fattoyev et al.(2018)]{Fattoyev18} Fattoyev, F. J., Piekarewicz, J., \& Horowitz, C. J. 2018, \prl, 120, 172702
\bibitem[Gamba et al.(2019)]{Gamba19} Gamba, R., Read, J. S., \& Wade, L. E. 2019, arXiv:1902.04616
\bibitem[Gendreau et al.(2012)]{Gendreau12} Gendreau K. C., Arzoumanian Z., Okajima T. {\it The Neutron star Interior Composition ExploreR (NICER): An Explorer mission of opportunity for soft x-ray timing spectroscopy[J]}. Proceedings of SPIE - The International Society for Optical Engineering, 2012, 8443:13.
\bibitem[Guillot et al.(2013)]{Guillot13} Guillot, S., Servillat, M., Webb, N. A., \& Rutledge, R. E. 2013, \apj, 772, 7
\bibitem[Guillot \& Rutledge(2014)]{Guillot14} Guillot, S. \& Rutledge, R. E. 2014, \apjl, 796, L3
\bibitem[Guillot(2016)]{Guillot16} Guillot S.¡¯s talk in Nusym2016, http://info.phys.tsinghua.edu.cn/enpg/nusym16/htmls/Program.htm.
\bibitem[Heinke(2004)]{Heinke04} Heinke, C. O. PhD thesis, Harvard University
\bibitem[Hinderer(2008)]{Hinderer08} Hinderer, T. 2008, \apj, 677, 1216
\bibitem[Hinderer et al.(2010)]{Hinderer10} Hinderer, T., Lackey, B. D., Lang, R. N., \& Read, J. S. 2010, \prd, 81, 123016
\bibitem[Kim et al.(2018)]{Kim18} Kim, Y. M., Lim, Y., Kwak, K., Hyun, C. H., \& Lee, C. H. 2018, \prc, 98, 065805
\bibitem[Kubis(2004)]{Kubis04} Kubis, S. 2004, \prc, 70, 065804
\bibitem[Kubis(2007)]{Kubis07} Kubis, S. 2007, \prc, 76, 025801
\bibitem[Lattimer \& Prakash(2000)]{Lattimer00} Lattimer, J. M., \& Prakash, M. 2000, PhR, 333, 121
\bibitem[Lattimer \& Prakash(2001)]{Lattimer01} Lattimer, J. M., \& Prakash, M. 2001, \apj, 550, 426
\bibitem[Lattimer \& Prakash(2007)]{Lattimer07} Lattimer, J. M., \& Prakash, M. 2007, \physrep, 442, 109
\bibitem[Lattimer \& Steiner(2014)]{Lattimer14} Lattimer, J. M., \& Steiner, A. W. 2014, Eur. Phys. J. A, 50, 40
\bibitem[Li \& Steiner(2006)]{Li06} Li, B. A., \& Steiner, A. W. 2006, Phys. Lett. B, 642, 436
\bibitem[Li \& Han(2013)]{Li13} Li, B. A., \& Han, X. 2013, Phys. Lett. B, 727, 276
\bibitem[Li(2017)]{Li17} Li, B. A. 2017, Nuclear Physics News, 27, 7
\bibitem[Li et al.(2019)]{Li19} Li, B. A., Krastev, P. G., Wen, D. H., \& Zhang, N. B. 2019, Eur. Phys. J. A, 55, 117
\bibitem[Lim \& Holt(2018)]{Lim18} Lim, Y., \& Holt, J. W. 2018, \prl, 121, 062701
\bibitem[Louren\c{c}o et al.(2019)]{Lourenco18} Louren\c{c}o, O., Dutra, M., Lenzi, C. H., Flores, C. V., \& Menezes, D. P. 2019, \prc, 99, 045202
\bibitem[Louren\c{c}o et al.(2019b)]{Lourenco19} Louren\c{c}o, O., Dutra, M., Lenzi, C. H., Biswa, S. K., Bhuyan, M., \& Menezes, D. P. 2019b, arXiv:1901.04529
\bibitem[Malik et al.(2018)]{Malik18} Malik, T., Alam, N., Fortin, M., Provid\^{e}ncia, C., Agrawal, B. K., Jha, T. K., Kumar, B., \& Patra, S. K. 2018, \prc, 98, 035804
\bibitem[Margalit \& Metzger(2017)]{Margalit17} Margalit, B., \& Metzger, B. D. 2017, \apj, 850, L19
\bibitem[Mereghetti(2011)]{Mereghetti11} Mereghetti, S. 2011, Astrophys. Space Sci. Proc., 21, 345
\bibitem[Miller(2013)]{Miller13} Miller, M. C. arXiv:1312.0029
\bibitem[Miller \& Lamb(2016)]{Miller16} Miller, M. C., \& Lamb, F. K. 2016, Eur. Phys. J. A, 52, 63
\bibitem[Most et al.(2018)]{Most18} Most, E. R., Weih, L. R., Rezzolla, L., \& Schaffner-Bielich, J. 2018, \prl, 120, 261103
\bibitem[Nandi et al.(2019)]{Nandi18} Nandi, R., Char, P., \& Pal, S. 2019, \prc, 99, 052802
\bibitem[N\"{a}ttil\"{a} et al.(2018)]{Nattla17} N\"{a}ttil\"{a}, J., Miller, M. C., Steiner, A. W., Kajava, J. J. E., Suleimanov, V. F., \& Poutanen, J. 2017, A\&A, 608, A31
\bibitem[Negele \& Vautherin(1973)]{Negele73} Negele, J. W., \& Vautherin, D. 1973, \nphysa, 207, 298
\bibitem[Oertel et al.(2017)]{Oertel17} Oertel, M., Hempel, M., Kl\"{a}hn, T., \& Typel, S. 2017, Rev. Mod. Phys., 89, 015007
\bibitem[Oppenheimer \& Volkoff(1939)]{Oppenheimer39} Oppenheimer, J., \& Volkoff, G. 1939, Phys. Rev., 55, 374
\bibitem[Ozel(2013)]{Ozel13} \"{O}zel, F. 2013, Rep. Prog. Phys., 76, 016901
\bibitem[Ozel et al.(2016)]{Ozel16} \"{O}zel, F., Psaltis, D.,  G\"{u}ver, T., Baym, G., Heinke, C., \& Guillot, S. 2016, \apj, 820, 28
\bibitem[Piekarewicz(2010)]{Piekarewicz10} Piekarewicz, J. 2010, J. Phys. G, 37, 064038
\bibitem[Piekarewicz(2018)]{Piekarewicz18b} Piekarewicz, J. Proceedings of the XIV International Workshop on Hadron Physics, Florian\'{o}polis, 18-23 March 2018 pages 12-37, arXiv:1805.04780
\bibitem[Piekarewicz \& Fattoyev(2019)]{Piekarewicz18} Piekarewicz, J., \& Fattoyev, F. J. 2019, \prc, 99, 045802
\bibitem[Radice \& Dai(2019)]{Radice19} Radice, D., \& Dai, L. 2019, Eur. Phys. J. A, 55, 50
\bibitem[Raithel et al.(2018)]{Raithel18} Raithel, C. A., \"{O}zel, F., \& Psaltis, D. 2018, \apjl, 857, L23
\bibitem[Raithel(2019)]{Raithel19} Raithel, C. A. 2019, Eur. Phys. J. A, 55, 80
\bibitem[Rezzolla et al.(2018)]{Rezzolla18} Rezzolla, L., Most, E. R., \& Weih, L. R. 2018, \apj, 852, L25
\bibitem[Ruiz et al.(2018)]{Ruiz18} Ruiz, M., Shapiro, S. L., \& Tsokaros, A. 2018, \prd, 97, 021501(R)
\bibitem[Shaw et al.(2018)]{Shaw18} Shaw, A. W., et al. 2018, \mnras, 476, 4713
\bibitem[Shibata et al.(2017)]{Shibata17} Shibata, M., Fujibayashi, S., Hotokezaka, K., Kiuchi, K., Kyutoku, K., Sekiguchi, Y., \& Tanaka, M. 2017, \prd, 96, 123012
\bibitem[Shibata et al.(2019)]{Shibata19} Shibata, M., Zhou, E., Kiuchi, K., \& Fujibayashi, S. 2019, \prd, 100, 023015
\bibitem[Shlomo et al.(2006)]{Shlomo06} Shlomo, S., Kolomietz, V. M., \& Col\`{o} G. 2006, Eur. Phys. J. A, 30, 23
\bibitem[Steiner et al.(2013)]{Steiner13} Steiner, A. W., Lattimer, J. M., \& Brown, E. F. 2013, \apjl, 765 L5
\bibitem[Steiner et al.(2018)]{Steiner18} Steiner, A. W., et al., \mnras, 476, 421
\bibitem[Suleimanov et al.(2016)]{Suleimanov16} Suleimanov, V. F.,Poutanen, J., Klochkov, D., \& Werner, K. 2016, Eur. Phys. J. A, 52, 20
\bibitem[Tews et al.(2017)]{Tews17} Tews, I., Lattimer, J. M., Ohnishi, A., \& Kolomeitsev, E. E. 2017, \apj, 848, 105
\bibitem[Tews et al.(2018)]{Tews18} Tews, I., Margueron, J., \& Reddy, S. 2018, \prc, 98, 045804
\bibitem[Tolman(1934)]{Tolman34} Tolman, R. C. 1934, Proc. Natl. Acad. Sci. U.S.A., 20, 3
\bibitem[Tong et al.(2019)]{Tong19} Tong, H., Zhao, P. W., Meng, J. 2019, arXiv:1903.05938
\bibitem[Tsang et al.(2019)]{Tsang19} Tsang, C. Y., Tsang, M. B., Pawel Danielewics, \& Lynch, W. G. 2019, arXiv:1905.02601
\bibitem[Xu et al.(2009)]{Xu09} Xu, J, Chen, L. W., Li, B. A., \& Ma, H. R. 2009, \apj, 697, 1549
\bibitem[Zhang et al.(2017)]{Zhang17} Zhang, N. B., Cai, B. J., Li, B. A., Newton, W. G., \& Xu, J. 2017, Nucl. Sci. Tech., 28, 181
\bibitem[Zhang et al.(2018)]{Zhang18} Zhang, N. B., Li, B. A. \& Xu, J. 2018, \apj, 859, 90
\bibitem[Zhang \& Li(2019)]{Zhang19} Zhang, N. B., \& Li, B. A. 2019, Eur. Phys. J. A, 55, 39
\bibitem[Zhang \& Li(2019b)]{Zhang19b} Zhang, N. B., \& Li, B. A. 2019b, J. Phys. G, 46, 014002
\bibitem[Zhang \& Li(2019c)]{Zhang19c} Zhang, N. B., \& Li, B. A. 2019c, \apj, 879, 99
\bibitem[Zhao et al.(2018)]{Zhao18} Zhao, T. Q., \& Lattimer, J. M. 2018 \prd, 98, 063020
\bibitem[Zhou et al.(2018)]{Zhou18} Zhou, E. P., Zhou, X., \& Li, A. 2018, \prd, 97, 083015
\bibitem[Zhou et al.(2019)]{Zhou19} Zhou, Y., Chen, L. W., \& Zhang, Z. 2019, \prd, 99, 121301
\end{thebibliography}
\end{document}